\documentclass{ws-ijmpc_rev}

\begin{document}

\markboth{Hiroshi Koibuchi}
{Shape transformations of a model of self-avoiding triangulated surfaces}


\title{Shape transformations of a model of self-avoiding triangulated surfaces of sphere topology
}

\author{Hiroshi Koibuchi
}

\address{Department of Mechanical and Systems Engineering, Ibaraki National College of Technology, Nakane 866 Hitachinaka, Ibaraki 312-8508, Japan
\\ koibuchi@mech.ibaraki-ct.ac.jp }



\maketitle


\begin{abstract}
We study a surface model with a self-avoiding (SA) interaction using the canonical Monte Carlo simulation technique on fixed-connectivity (FC) triangulated lattices of sphere topology. The model is defined by an area energy, a deficit angle energy, and the SA potential. A pressure term is also included in the Hamiltonian. The volume enclosed by the surface is well defined because of the self-avoidance. We focus on whether or not the interaction influences the phase structure of the FC model under two different conditions of pressure ${\it \Delta} p$; zero and small negative. The results are compared with the previous results of the self-intersecting model, which has a rich variety of phases; the smooth spherical phase, the tubular phase, the linear phase, and the collapsed phase. We find that the influence of the SA interaction on the multitude of phases is almost negligible except for the evidence that no crumpled surface appears under ${\it \Delta} p\!=\!0$ at least even in the limit of zero bending rigidity $\alpha\!\to \!0$. The Hausdorff dimension is obtained in the limit of $\alpha\!\to \!0$ and compared with previous results of SA models, which are different from the one in this paper. 

\end{abstract}

\ccode{PACS Nos.: 11.25.-w,  64.60.-i, 68.60.-p, 87.10.-e, 87.15.ak}

\section{Introduction}\label{intro}
Over the past few decades, a considerable number of studies have been conducted on the surface models. The model was constructed for strings and membranes \cite{NELSON-SMMS2004,Gompper-Schick-PTC-1994,WIESE-PTCP19-2000,Bowick-PREP2001,SEIFERT-LECTURE2004,GOMPPER-SMMS2004,WHEATER-JP1994}, and it was defined on the basis of the differential geometric notion of curvatures \cite{HELFRICH-1973,POLYAKOV-PLB1981,POLYAKOV-NPB1986,KLEINERT-PLB1986}. The so-called crumpling transition is a shape transformation between the smooth phase at sufficiently large bending rigidity $\alpha$ and the collapsed phase at $\alpha\!\to\!0$, and it has long been studied both theoretically \cite{Peliti-Leibler-PRL1985,PKN-PRL1988,DavidGuitter-EPL1988,Kownacki-Mouhanna-2009PRE} and numerically \cite{BILLOIRE-DAVID-NPB1986,BKKM-NPB1986,KANTOR-NELSON-PRA1987,Baum-Ho-PRA1990,CATTERALL-NPBSUP1991,AMBJORN-NPB1993}. While the transition is considered as a continuous one \cite{DavidGuitter-EPL1988,Kownacki-Mouhanna-2009PRE}, a possibility that it is of first-order is pointed out \cite{PKN-PRL1988}, and renormalization group studies \cite{NISHIYAMA-PRE-2004} and recent numerical studies \cite{KD-PRE2002,KOIB-PRE-2004,KOIB-PRE-2005,KOIB-NPB-2006} predict that the transition is of first order. The transition was observed in the canonical surface model on relatively large sized surfaces \cite{KOIB-PRE-2005}.

 In addition to the smooth and the collapsed phases, a variety of phases including the tubular phase are observed in surface models \cite{KOIB-PRE-2007}, which are defined by a one-dimensional bending energy on the cytoskeletal structure. A planar phase and an oblong linear phase can be seen in a model \cite{KOIB-EPJB-2007-3}, which is defined by a one-dimensional bending energy and the Nambu-Goto area energy on the fixed-connectivity (FC) surface. A surface model defined by a deficit angle energy also has a rich variety of phases including a tubular phase \cite{KOIB-PRE-2004-2}. It must be noted that these phase transitions can be observed on relatively smaller surfaces in contrast to the above mentioned crumpling transition of the canonical surface model.
 
To construct a surface model, the self-avoiding (SA) property should be taken into account if we focus on membranes \cite{KANTOR-KARDAR-NELSON-PRL1986,KANTOR-KARDAR-NELSON-PRA1987,PLISCHKE-BOAL-PRA1988,Ho-Baum-EPL1990,BAUM-JPIF1991,BAUM-RENZ-EPL1991,GREST-JPIF1991,Gompper-Kroll-JPF1993,Gompper-Kroll-PRE1995,Munkel-Heermann-PRL1995,BCTT-PRL2001,BOWICK-TRAVESSET-EPJE2001}. However,  numerical studies of the SA surfaces are very time consuming because of the non-local property of the interactions. The simulations on such large surfaces like those in {Refs.} \cite{KOIB-PRE-2005,KOIB-NPB-2006} are still not feasible on currently available computers. Nevertheless, the numerical studies on the above mentioned variety of phases in those specific models are considered to be feasible on the SA surfaces.

Therefore, it is very interesting to study whether or not the SA interaction influences the phase structure in those models without SA interactions (phantom surface models). It is possible that the multitude of phases is strongly influenced by the SA interactions. In fact, no completely-collapsed phase is observed in FC SA surfaces \cite{PLISCHKE-BOAL-PRA1988,Ho-Baum-EPL1990,BAUM-JPIF1991,BAUM-RENZ-EPL1991,GREST-JPIF1991,Gompper-Kroll-JPF1993,BOWICK-TRAVESSET-EPJE2001}.   Moreover, the SA interaction is expected to play a non-trivial role in the membrane morphology even at the smooth phase. It was recently reported that SA property is essential for a variety of shapes of the so-called excess cone at high bending regime \cite{SWAMM-PRL-2010}.   

In this paper, we study the surface model in Ref. \cite{KOIB-PRE-2004-2} with a SA interaction on FC triangulated surfaces by using the canonical Monte Carlo (MC) simulation technique.
The smooth spherical phase, the tubular phase, the linear phase, and the collapsed phase are seen in the FC phantom surface model \cite{KOIB-PRE-2004-2}. Our interests are focused on whether or not such a variety of phases, including the collapsed phase, are influenced by the SA interaction. Two different values of pressure ${\it \Delta}p$ are assumed such that ${\it \Delta}p$ is zero and small negative.  

This paper is organized as follows: in Section \ref{triangulated_models}, we make a brief outline of the current results of the numerical studies of phantom surface models and SA surface models on triangulated surfaces. In Section \ref{model}, we define the model with a SA interaction, which is slightly different from the currently well-known SA interactions for numerical studies. The Monte Carlo simulation technique is shown in Section \ref{MC-Techniques}, and the numerical results are presented in Section \ref{Results}. We summarize the results in the final Section \ref{Conclusion}.  

\section{Triangulated surface models}\label{triangulated_models}
\subsection{Phase structure of phantom surface models}\label{current_results}
In this subsection, we give a brief outline of the phantom surface models on triangulated lattices in ${\bf R}^3$ and the current numerical results. We start with the continuous model, which is given by the continuous Hamiltonian $S\!=\!S_1\!+\!\alpha S_2$, where $S_1\!=\!\int \sqrt{g}d^2x g^{ab} \partial_a X^\mu \partial_b X^\mu$ and $S_2\!=\!(1/2)\int \sqrt{g}d^2x g^{ab} \partial_a n^\mu \partial_b n^\mu$. $S_1$ is just identical with the action of Polyakov string \cite{POLYAKOV-PLB1981,POLYAKOV-NPB1986}, where $X^\mu$ denotes a mapping from a two-dimensional surface $M$ to ${\bf R}^3$ and represents the surface position in ${\bf R}^3$, $g^{ab}$ is the inverse of the metric tensor $g_{ab}$ of $M$, and $g$ is the determinant of $g_{ab}$. The variables $(x_1,x_2)$ represent a local coordinate of $M$. The image $X(M) (\subset {\bf R}^3)$ is the surface, which is triangulated in numerical studies. The symbol $n^\mu$ in $S_2$ denotes a unit normal vector of $X(M)$, and $S_2$ is called the bending energy, and $\alpha$ is the bending rigidity.

If $g_{ab}$ is fixed to the Euclidean metric $\delta_{ab}$ and $X(M)$ is triangulated by piecewise linear triangles, then we have
$S\!\!=S_1\!+\!\alpha S_2$, $S_1\!=\!\sum_{ij}\left(X_i\!-\!X_j \right)^2$, $S_2\!=\!\sum_{ij}\left(1\!-\! {\bf n}_i\cdot{\bf n}_j\right)$, where $X_i (\in {\bf R}^3)$ in $S_1$ is the position of the vertex $i$, ${\bf n}_i$ in $S_2$ is a unit normal vector of the triangle $i$. The FC model is statistical mechanically defined by the partition function 
\begin{equation} 
\label{Part-Func}
 Z_{\rm fix} = \int^\prime \prod _{i=1}^{N} d X_i \exp\left[-S(X)\right], \qquad({\rm fixed}), 
\end{equation}
where the prime in $\int^\prime  \prod _{i=1}^{N}d X_i$ denotes that the three-dimensional multiple integrations are performed by fixing the center of mass of the surface at the origin of ${\bf R}^3$ to remove the translational zero mode.  We call the FC model defined by the energies $S_1$ and $S_2$ as the "canonical" surface model. It was reported that the canonical model on surfaces of sphere topology undergoes a first-order transition at finite $\alpha_c$ between the smooth phase at $\alpha\to\infty$ and the collapsed phase at $\alpha\to 0$ \cite{KOIB-PRE-2005}. The role of the Gaussian bond potential $S_1$ is to make the mean bond length constant and, hence, $S_1$ can be replaced by a Lennard-Jones type potential \cite{KD-PRE2002} and also by a hard-wall potential \cite{KOIB-NPB-2006}.  In a surface model on triangulated lattices of the seminal paper {Ref.} \cite{KANTOR-NELSON-PRA1987} of Kantor and Nelson, $S_1$ is given by a hard-core and hard-wall potential. { This type of potential can be used as a SA potential, which is described in the following sebsection. }
 
A variation of the canonical model is obtained by replacing $S_1$ with the Nambu-Goto area energy $S_{\it \Delta}=\sum_{\it \Delta}A_{\it \Delta}$, where $A_{\it \Delta}$ is the area of the triangle ${\it \Delta}$. $S_{\it \Delta}$ is also obtained from the above mentioned continuous Hamiltonian $S_1$ by fixing $g_{ab}$ as the induced metric $g_{ab}\!=\!{\partial_a X^\mu}{\partial_b X^\mu}$ of the mapping $X$. We call a model as the Nambu-Goto surface model if the Hamiltonian includes $S_{\it \Delta}$ as the bond potential term. It is well-known that the Nambu-Goto model with the canonical bending energy $S_2\!=\!\sum_{ij}\left(1\!-\! {\bf n}_i\cdot{\bf n}_j\right)$  is ill-defined in the sense that no equilibrium configuration is obtained in the numerical simulations \cite{ADF-NPB-1985}. The  ill-definedness comes from the fact that the area $A_{\it \Delta}$ is totally independent of the shape of ${\it \Delta}$, and the oblong and very thin triangles, which are considered as singular triangles, dominate the surface configurations in the whole range of $\alpha$. However, if the canonical bending energy $S_2$ is replaced by a deficit angle energy such as $S_2^{\rm int}\!=\!\sum_i \left(\delta_i-\phi_0\right)^2$ or $S_2^{\rm int}\!=\!\sum_i |\delta_i-\phi_0|$  \cite{Baillie-Johnston,BEJ,BIJJ}, the model turns to be well defined except in the limit of $\alpha\to 0$ \cite{KOIB-PRE-2004-2}. The symbol $\delta_i$ in $S_2^{\rm int}$ is the sum of internal angles of triangles meeting at the vertex $i$, and $\phi_0$ is a constant and fixed to $\phi_0\!=\!2\pi$ if the surface is closed. The deficit angle energy $S_2^{\rm int}\!=\!-\sum_i \log \left(\delta_i/2\pi\right)$ is possible on closed surfaces such as a sphere \cite{KOIB-PRE-2004-2,KOIB-EPJB-2004}. Those deficit angle energies are called as the intrinsic curvature energy. The reason of the variety of phases in the Nambu-Goto model with $S_2^{\rm int}$  \cite{KOIB-PRE-2004-2} seems that both $S_{\it \Delta}$ and $S_2^{\rm int}$ are insensitive to the surface shape. In fact, $S_{\it \Delta}$ is independent of whether or not the surface is composed of almost-regular triangles or oblong triangles. $S_2^{\rm int}$ is also independent of whether the surface is planar or cylindrical. Nevertheless, both of the smooth phase and the collapsed phase are stable on the disk surface \cite{KOIB-PLA-2005} and on the torus \cite{KOIB-PLA-2006-1}. {In this, paper we study the Nambu-Goto surface model with the intrinsic curvature {$S_2^{\rm int}$}. }

{We should comment on the reason why we use {$S_2^{\rm int}\!=\!-\sum_i \log \left(\delta_i/2\pi\right)$} as the intrinsic curvature energy. The origin of {$S_2^{\rm int}\!=\!-\sum_i \log \left(\delta_i/2\pi\right)$} is the measure factor {$q_i^\sigma$} in the integrations {$\int^\prime \prod _{i=1}^{N} d X_i q_i^\sigma$} in $Z$, where {$q_i$} is the coordination number of the vertex {$i$} and {$\sigma (\!=\!3/2)$} is a constant \cite{FDAVID-NPB-1985}. By identifying {$q_i$} with {$\delta_i$} and extending the constant {$\sigma$} to the variable coefficient {$\alpha$}, we  have the expression {$-\alpha\sum_i \log \left(\delta_i\right)$}. Including the normalization factor {$2\pi$}, we have the curvature energy {$S_2^{\rm int}\!=\!-\sum_i \log \left(\delta_i/2\pi\right)$}. }

{We should also comment on the fact that the mean value of {$S_1$} is constant such that {$\langle S_1/N\rangle \!=\!3/2$} even in the limit of {$b\to 0$}. The reason for  {$\langle S_1/N\rangle \!=\!3/2$} is understood from the scale invariant property of the partition function \cite{WHEATER-JP1994}. In fact, by rescaling the integration variable in {$Z$} such that {$X\to \lambda X$}, we obtain {$Z(\lambda)\!=\!\lambda^{3(N\!-\!1)}\int^\prime \prod_{i=1}^N d X_i \exp\left[-S(\lambda X) \right]$}, where {$S(\lambda X)\!=\!\lambda^2 S_1\!+\!bS_2$}. The scale invariance of {$Z$} indicates that {$Z(\lambda)$} is independent of {$\lambda$} and, therefore, is represented by {$\partial Z(\lambda)/\partial \lambda |_{\lambda=1} \!=\!0$}. Thus, we have {$\langle S_1/N\rangle\!=\!3(N\!-\!1)/2N\!\simeq\!3/2$}.  }

A variety of phases can also be seen in a model, which is obtained by replacing $S_1$ and $S_2$ of the canonical model with $S_{\it \Delta}$ and the one-dimensional bending energy $S_2^{\rm 1\!-\!d}$, respectively \cite{KOIB-EPJB-2007-3}. In this case, $S_2^{\rm 1\!-\!d}$ is sensitive to the surface shape, while $S_{\it \Delta}$ is not as mentioned above. 

A variation of the canonical model is obtained also by including fluidity, which represents a lateral diffusion of vertices \cite{GOMPPER-SMMS2004,BILLOIRE-DAVID-NPB1986,BKKM-NPB1986,Baum-Ho-PRA1990,CATTERALL-NPBSUP1991,AMBJORN-NPB1993}. This two-dimensional fluidity is defined on dynamically triangulated surfaces, where the triangulation ${\cal T}$ is considered as a dynamical variable of the model. The partition function of the model with fluidity is thus given by 
\begin{equation} 
\label{Fluid_Part-Func}
Z_{\rm flu} \!=\! \sum _{\cal T}\int^\prime \prod _{i=1}^{N} d X_i \exp\left[-S(X, {\cal T})\right], \qquad({\rm fluid})
\end{equation}
 where $S(X, {\cal T})$ represents that $S$ is dependent on the variables $X$ and ${\cal T}$, and $\sum _{\cal T} $ represents the sum over all possible triangulations. In the fluid model corresponding to the canonical model, we cannot see the transition, which is seen in the canonical model on FC spherical surfaces. This is expected from the phase structure of compartmentalized fluid surfaces \cite{KOIB-EPJB-2007-1}, where the lateral diffusion is allowed only inside the compartment, which is a sublattice structure on the surface. In this compartmentalized model, a first-order transition, which is considered to be identical to the one in the canonical FC model, disappears if the compartment size $L_C$ is increased. The homogeneous fluid surface is obtained from the compartmentalized fluid surface by maximizing $L_C$ such that the surface is composed of a single compartment or the surface has no compartment. Thus, we understand that the transition, which is observed on the compartmentalized fluid surfaces at relatively small $L_C$, cannot be observed on the homogeneous fluid surfaces. The Nambu-Goto model with the intrinsic curvature energy is well-defined even on the fluid surfaces and has a variety of phases \cite{KOIB-EPJB-2007-2}.    

By combining two different sets of ball-spring systems, Boal and Seifert introduced a fluid surface model with cytoskeletal structures, which is a two-components network model for red cells \cite{BOAL-ZEIFERT-PRL1992}. If a curvature energy is introduced on the compartment in place of the canonical bending energy $S_2$ in the compartmentalized fluid surface model in {Ref.} \cite{KOIB-EPJB-2007-1}, we have also fluid surface models with cytoskeletal structures \cite{KOIB-PRE-2007}. A large variety of shape transformations are observed in such inhomogeneous fluid surface models, where the bond potential $S_1$ is  the Gaussian bond potential, and the curvature energy $S_2$ is the one-dimensional bending energy $S_2^{\rm 1\!-\!d}$ defined only on the compartments, which are one-dimensional objects linked with junctions \cite{KOIB-PRE-2007}. The phase structure depends on the elasticity at the junctions; a planar phase, and a tubular phase are observed in those models. The reason for such a variety of phases is closely connected to the cytoskeletal structure and the lateral diffusion of vertices. In fact, $S_2^{\rm 1\!-\!d}$ is considered to be insensitive to the surface shape, because $S_2^{\rm 1\!-\!d}$ is defined only on the compartments in contrast to the model in Ref. \cite{KOIB-EPJB-2007-3}, where $S_2^{\rm 1\!-\!d}$ is defined all over the lattice. The surface shape is not always uniquely determined if the curvature is given only at small part of the surface, and moreover large surface fluctuations are expected in the compartmentalized model in Ref. \cite{KOIB-PRE-2007} due to the lateral diffusion of vertices inside the compartments. 

\subsection{Self-avoiding surface models}\label{SA_results}
The current studies that have been conducted on SA surfaces are considered to be still in the pioneer stage. In this subsection, we briefly comment on the existing SA surface models and the results of the numerical studies. The SA surface model is defined by a SA interaction, which is an extension of the Hamiltonian of the Edward model for polymers \cite{WIESE-PTCP19-2000,Bowick-PREP2001}. The phase structure of the SA models has been extensively studied \cite{KANTOR-KARDAR-NELSON-PRL1986,KANTOR-KARDAR-NELSON-PRA1987,PLISCHKE-BOAL-PRA1988,Ho-Baum-EPL1990,BAUM-JPIF1991,BAUM-RENZ-EPL1991,GREST-JPIF1991,Gompper-Kroll-JPF1993,Gompper-Kroll-PRE1995,Munkel-Heermann-PRL1995,BCTT-PRL2001,BOWICK-TRAVESSET-EPJE2001}, although the total number of studies are currently considered to be far smaller than those of the phantom surfaces. 

We have two types of SA models for numerical studies: the ball-spring (BS) model and the impenetrable plaquette (IP) model. The BS model is defined on two-dimensional networks, which are composed of vertices and bonds connecting two nearest neighbor vertices by a hard-core and hard-wall potential \cite{KANTOR-KARDAR-NELSON-PRL1986,KANTOR-KARDAR-NELSON-PRA1987}. The size of ball as the vertices and the length of spring as bonds are constrained such that no vertex can move from one side of a triangle to the other side. This SA potential of the BS model is defined between all pairs of vertices, however, the simulations are slightly less time-consuming than those of the IP model. The SA interaction of the IP model is defined such that the triangles are constrained to avoid intersecting. Although the simulations of the IP model are relatively time consuming, the IP model seems advantageous to the BS model. In fact, two neighboring triangles $i$ and $j$ of the IP model can completely bend such that $1\!-\!{\bf n}_i\cdot{\bf n}_j\!=\!2$, while in the BS model the bending angle $\theta_{ij}$ is constrained such that $\theta_{ij}\!<\!\theta_0$, where $\cos \theta_{ij}\!=\!{\bf n}_i\cdot{\bf n}_j$, and $\theta_0(<\pi)$ is determined by the SA potential.     

The crumpling transition is reported to disappear from the SA FC surfaces \cite{PLISCHKE-BOAL-PRA1988,Ho-Baum-EPL1990,BAUM-JPIF1991,BAUM-RENZ-EPL1991,GREST-JPIF1991,Gompper-Kroll-JPF1993,BOWICK-TRAVESSET-EPJE2001}; this is because no completely-collapsed phase appears in the SA FC surfaces. In fact, no crumpled phase is observed in both of the BS model \cite{PLISCHKE-BOAL-PRA1988,Ho-Baum-EPL1990,GREST-JPIF1991} and the IP model \cite{BAUM-JPIF1991,BAUM-RENZ-EPL1991,Gompper-Kroll-JPF1993,BOWICK-TRAVESSET-EPJE2001}. To the contrary, the smooth phase is expected to remain unchanged from that of the phantom surfaces. However, numerical results are not always universal; in fact, the Hausdorff dimension of the IP model of \cite{Gompper-Kroll-JPF1993} is $H\!\simeq\!2.3$, while that of {Ref.} \cite{BOWICK-TRAVESSET-EPJE2001} is $H\!=\!2.1(1)$, although both models are defined on the disk surface.  {As mentioned in the Introduction, it is possible that the SA interaction plays a non-trivial role in membrane shapes in the smooth phase \cite{SWAMM-PRL-2010}.} Thus, we should study the SA model more extensively.

To summarize the comments including those in the previous subsection, we have several phantom surface models, which have a multitude of phases. The models are considered as non-trivial variations of the canonical surface model. The phase structures of almost all models have not yet been studied on the SA surfaces. The current understanding of the phase structure of SA surfaces are as follows: the crumpling transition disappears from the FC model, because the SA interaction prohibits the surfaces from collapsing in both of the BS model and the IP model. The smooth phase of the SA surfaces are considered to be {almost} identical to the smooth phase of the phantom surfaces{, while the membrane shapes are expected to be influenced by the SA interaction under some specific conditions.} 

\section{Model}\label{model}
In this section, we define a SA model, which corresponds to the phantom surface models in {Refs.} \cite{KOIB-PRE-2004-2,KOIB-EPJB-2007-2}. Triangulated lattices of sphere topology are assumed to define the model, and the lattices are constructed using the icosahedron. By splitting the edges and faces of the icosahedron, we have a lattice of size $N\!=\!10\ell^2\!+\!2$, where $\ell$ is the devision number of an edge of the icosahedron. The coordination number $q$ of vertices is $q\!=\!6$ almost everywhere excluding 12 vertices of $q\!=\!5$. The lattice is characterized by the three numbers $N$, $N_B(=\!3N\!-\!6\!=\!30\ell^2)$, and $N_T(=\!2N\!-\!4\!=\!20\ell^2)$, which are the total number of vertices, the total number of bonds, and the total number of triangles, respectively.
The lattices used in {Ref.} \cite{KOIB-PRE-2004-2} are random lattices, of which the coordination number is not always uniform and, they are slightly different from the lattices constructed as above. However, the phase structure of FC surface models is expected to be independent of the lattice structure \cite{KOIB-NPB-2006}. 

The dynamical variable of the FC model is the position $X_i(\in {\bf R}^3)$ of the vertex $i(=1,\cdots,N)$. The partition functions of the model are given by Eq. (\ref{Part-Func}). The Hamiltonian $S(X)$ is defined by a linear combination of the area energy $S_1$, a curvature energy $S_2$, the pressure term $-{\it \Delta}p \,V$, and a SA potential $U$, such that
\begin{eqnarray}
\label{Disc-Eneg} 
&& S(X)=S_1 + \alpha S_2 -{\it \Delta}p \,V+ U, \nonumber \\
&& S_1=\sum_{\it \Delta} A_{\it \Delta},  \quad S_2=-\sum_{i} \log \left(\delta _i/2\pi\right),  \\
&& U=\sum_{{\it \Delta},{\it \Delta}^\prime} U({\it \Delta},{\it \Delta}^\prime), \quad U({\it \Delta},{\it \Delta}^\prime)= \left\{
                       \begin{array}{@{\,}ll}
                          \infty & \; ({\rm triangles}\; {\it \Delta},{\it \Delta}^\prime\;{\rm intersect}  ), \\
                          0 & \; ({\rm otherwise}). 
                       \end{array} 
                        \right. \nonumber \\ \nonumber
\end{eqnarray} 
$S_1$ is the sum over the area $A_{\it \Delta}$ of triangle ${\it \Delta}$. The symbol $\delta_i$ in $S_2$ is the sum of internal angle of triangles meeting at the vertex $i$. $S_2$ can be called a deficit angle energy, although $S_2$ is different from the sum of the deficit angle $\delta_i\!-\!2\pi$ of the vertex $i$. If $S_2$ is defined without "$\log$" and is given by $\sum _i(\delta-2\pi)$, then $S_2$ depends only on the surface topology and is a constant on piece-wise linearly triangulated surfaces. However, $S_2$ in Eq. (\ref{Disc-Eneg}) is well-defined as a curvature energy because of the $\log$ function as mentioned in Section \ref{current_results}. The symbol $\alpha[kT]$ denotes the bending rigidity, where $k$ is the Boltzmann constant and $T$ is the temperature. 

$V$ is the volume enclosed by the surface, and ${\it \Delta}p$ is the pressure which is defined by ${\it \Delta}p\!=\!p_{\rm in}\!-\!p_{\rm out}$, where $p_{\rm out}$ ($p_{\rm in}$) is the pressure outside (inside) the surface. If $p_{\rm out}$ is assumed to be $p_{\rm out}\!=\!0$, then the positive (negative) ${\it \Delta}p$ implies that $p_{\rm in}$ is positive (negative). We should note also that the volume $V$ is well defined only if the surface is self-avoiding. $V$ is bounded below such that $V\geq 0$ in the SA surfaces, while $V$ can be negative in non SA surfaces. 

\begin{figure}[!h]
\centering
\includegraphics[width=12.5cm]{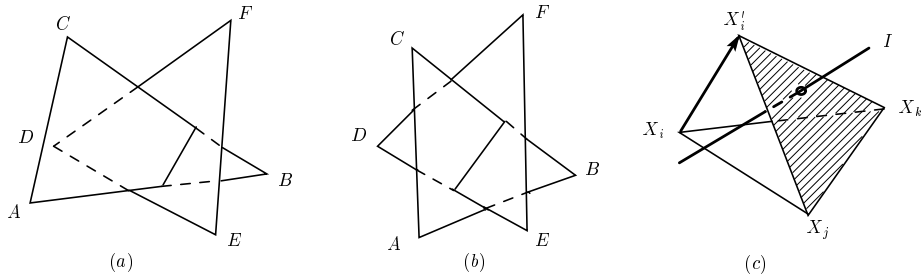}  
\caption{ (a),(b) Two intersecting triangles, and (c) an intersection of the bond $I$ and the triangles with the vertices $X_i^\prime$, $X_j$ and $X_k$, where $X_i^\prime$ is a new position of the vertex $i$. In (a), the bonds $AB$ and $BC$ of the triangle $ABC$ intersect with the triangle $DEF$, while no bond of the triangle $DEF$ intersects with the triangle $ABC$. In (b), the bond $BC$ of the triangle $ABC$ intersects with the triangle $DEF$, and the bond $DE$ of the triangle $DEF$ intersects with the triangle $ABC$. 
 } 
\label{fig-1}
\end{figure}
$\sum_{{\it \Delta},{\it \Delta}^\prime}$ in the SA potential $U$ denotes the sum over all pairs of non nearest neighbor triangles ${\it \Delta}$ and ${\it \Delta}^\prime$. The potential $U({\it \Delta},{\it \Delta}^\prime)$ is defined such that any pairs of non nearest neighbor triangles ${\it \Delta}$ and ${\it \Delta}^\prime$ should not be intersecting. {Figure {\ref{fig-1}}(a) shows} two pairs of intersecting triangles, in which the triangle $ABC$ penetrates the triangle $DEF$ or in other words the bonds $AB$ and $BC$ intersect with the triangle $DEF$. On the contrary in  {Fig. {\ref{fig-1}}(b)}, the triangle $ABC$ and the triangle $DEF$ intersects with each other, or in other words a bond of one triangle intersects with the other triangle and vise versa. We describe the numerical implementation of the SA interaction $U$ in detail in the following section.

The SA potential $U$ in Eq. (\ref{Disc-Eneg}) is not identical to the one assumed in the SA model in Ref. \cite{BOWICK-TRAVESSET-EPJE2001} and, hence, the surface is completely self-avoiding under the potential $U$. In fact, the triangles are allowed to intersect with finite energy in Ref. \cite{BOWICK-TRAVESSET-EPJE2001}, while those in the model of this paper are prohibited to intersect with each other because of the infinite energy assumed in $U$. 

Finally in this section, we comment on how to compute the volume $V$ enclosed by the surface. The initial value of $V$ in the simulations is assumed such that $V\!=\!4\pi r^3/3$, where $r$ is the radius of the initial configuration of sphere lattice. This initial value $V\!=\!4\pi r^3/3$ is slightly larger than the real volume, because the surface is linearly triangulated. However, it is almost evident that the deviation can be negligible in the limit of $N\!\to\!\infty$. The volume $V$ changes during the simulations according to the rule $V\to V+{\it \Delta}V$ every update of vertex, where ${\it \Delta}V$ is the volume of small tetrahedra, such as the one shown in  {Fig. {\ref{fig-1}}(c)}. ${\it \Delta}V$ is positive or negative, which is determined according to whether the new position $X_i^\prime$ is outside or inside the surface, in which the orientation is uniquely fixed by a normal vector of each triangle. We should note that ${\it \Delta}V$ is well defined only when the surface is self-avoiding. It is apparent that  ${\it \Delta}V$ is not well defined when some part of volume element of ${\it \Delta}V$ is shared by some other ${\it \Delta}V^\prime$, i.e., the surface is allowed to self intersect. 

{The enclosed volume {$V$} can also be computed by using the divergence theorem applying the position vector {${\bf r}_i$} of the center of mass of the triangle $i$. Not only {${\it \Delta}V$} but also {$V$} is exactly identical to the one obtained by the above mentioned technique. A very small deviation can be seen in the total volume {$V$}, however, it is less than {$1\%$} even in the cup like phase on the {$N\!=\!1442$} surface during the simulations. This small deviation of {$V$} is the one between $V\!=\!4\pi r^3/3$ and {$V$} of the initial triangulated sphere.}

\section{Monte Carlo technique}\label{MC-Techniques}
The canonical Metropolis Monte Carlo (MC) technique is employed for simulating the integrations of the variables $X$ in $Z_{\rm fix}$ of Eq. (\ref{Part-Func}). The three-dimensional random move $X\to X^\prime\!=\!X\!+\!\delta X$ is accepted with the probability ${\rm Min}[1,\exp (-\delta S)]$, where $\delta S$ is given by $\delta S\!=\!S({\rm new})\!-\!S({\rm old})$ under the constraint of the potential $U$. The symbol $\delta X$ is randomly chosen in a small sphere, whose radius is fixed in the simulations such that the acceptance rate $r_X$ of $X^\prime$ should be approximately $r_X\!=\!50\%$. 

The constraint of $U({\it \Delta},{\it \Delta}^\prime)$ in Eq. (\ref{Disc-Eneg}) is composed of two different constraints on a new vertex position as follows: let $X_i$ and $X_i^\prime$ denote the current position and the new position of the vertex $i$ as shown in Fig. \ref{fig-1}(c). The shaded triangle in Fig. \ref{fig-1}(c) forms a new surface.  One of the constraint imposed on $X_i^\prime$ is that the new triangle $i^\prime jk$ has no intersection with the disjoint bonds, where "disjoint bonds" are the edges of triangles disconnected with the triangle $ijk$. The other constraint is that every new bond, such as the bond $i^\prime j$ in Fig. \ref{fig-1}(c), has no intersection with the disjoint triangles, where "disjoint triangles" are those disconnected with the bond $ij$. These two constraints imposed on $X_i^\prime$ make the surface self-avoiding in the sense that any two disjoint triangles have no intersection with each other. 

The first constraint prohibits the new triangle $i^\prime jk$ shown in Fig. \ref{fig-1}(c) from being penetrated by disjoint triangles. The second constraint imposed on $X_i^\prime$ prohibits the new triangle  $i^\prime jk$ from penetrating some other triangles. The intersection of the triangles shown in Fig. \ref{fig-1}(b) is prohibited by both of the constraints, while the intersection in Fig. \ref{fig-1}(a) is prohibited only by one constraint and is not prohibited by the other constraint. This is the reason why two constraints are necessary to make the surface self-avoiding by checking an intersection of a bond and a triangle. 

We assume a sphere of radius $R_0$ at the center of mass of the triangle $i^\prime jk$ shown in Fig. \ref{fig-1}(c), and check whether or not the triangle intersects with disjoint bonds inside the sphere. The check of intersection in the second constraint is also performed assuming the sphere of size $R_0$ at the center of the bond $i^\prime j$. The radius $R_0$ is assumed to be $R_0\!=\!6\langle L\rangle$, where $\langle L\rangle$ is the mean bond length. As a consequence, the computational time is reduced by $20\%\sim 60\%$ or more, which depends on $\alpha$. 

The bond length $L$ and the triangle area $A_{\it \Delta}$ are bounded below such that $L>1\!\times\! 10^{-7}$ and $A_{\it \Delta}>0.5\!\times\! 10^{-7}$ in the simulations. The final results of the simulations are considered to be independent of these lower bounds, because these bounds are sufficiently small and almost all bond lengths and triangle areas are larger than these values.

The total number of MC sweeps (MCS) after the thermalization MCS is about $1\!\times\!10^7\sim 2\!\times\! 10^7$ on the $N\!=\!1442$ surface, and relatively small number of MCS is assumed on the smaller surfaces. The total number of the thermalization MCS is about $0.5\!\times\! 10^6$. The thermalization MCS in the collapsed tubular phase is very large; it is sometimes $1\!\times\! 10^7$ or more at the phase boundary close to the cup like phase on the $N\!=\!1442$ surface. Intersection of bonds with triangles is checked every $5\!\times\!10^5$ MCS throughout the simulation; the check is performed between every disjoint pair of bond and triangle.  No intersection is observed at every assumed value of $\alpha$ including $\alpha\!=\!0$ and ${\it \Delta}p$. 

\section{Results}\label{Results}

\begin{figure}[htb]
\centering
\includegraphics[width=12.5cm]{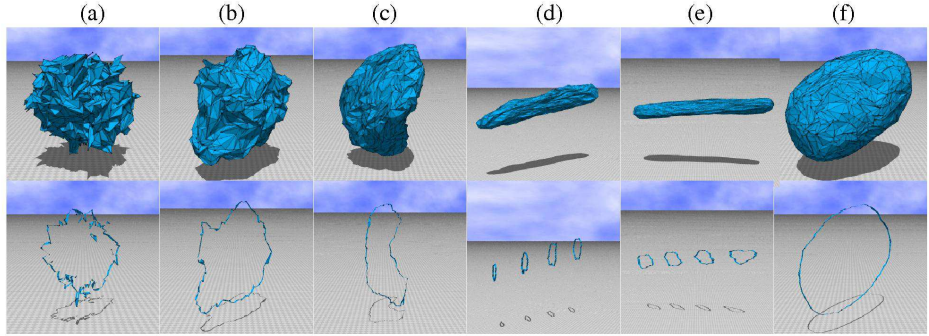}  
\caption{The snapshots of FC surfaces and the surface sections of size $N\!=\!1442$ obtained under ${\it \Delta}p\!=\!0$ at (a) $\alpha\!=\!0$ (wrinkled), (b) $\alpha\!=\!100$ (wrinkled), (c) $\alpha\!=\!500$ (wrinkled), (d) $\alpha\!=\!1000$ (tubular), (e) $\alpha\!=\!1\!\times\!10^4$ (tubular), and (f) $\alpha\!=\!2\!\times\!10^4$ (smooth spherical).} 
\label{fig-2}
\end{figure}
The snapshots of FC surfaces at ${\it \Delta}p\!=\!0$ are shown in Figs. \ref{fig-2}(a)--\ref{fig-2}(f). The surface size is $N\!=\!1442$. The assumed bending rigidities are (a) $\alpha\!=\!0$, (b) $\alpha\!=\!100$, (c) $\alpha\!=\!500$, (d) $\alpha\!=\!1000$, (e) $\alpha\!=\!1\!\times\!10^4$, and (f) $\alpha\!=\!2\!\times\!10^4$. The scales of the figures are all different from each other. The surfaces shown in the figure are considered to be in (a),(b),(c) the wrinkled phase, (d),(e) the tubular phase, and (f) the smooth spherical phase. The surface in Fig. \ref{fig-2}(a) can be called a collapsed surface because the surface is highly fluctuating, however, it encloses empty space inside the surface and, therefore, the surface is not always crumpled in the limit of $\alpha\!\to\! 0$. The spherical surfaces in Figs. \ref{fig-2}(b) and \ref{fig-2}(c) look slightly smooth, however, they are apparently different from the surface at sufficiently large $\alpha$ shown in Fig. \ref{fig-2}(f). The surfaces in  Figs. \ref{fig-2}(d) and \ref{fig-2}(e) can be called a tubular surface. The surface in Fig. \ref{fig-2}(f) is very smooth and can be called the smooth spherical surface. All of the phases, excluding the wrinkled phase, correspond to those of the same model without the SA interaction in {Ref.} \cite{KOIB-PRE-2004-2}. The collapsed phase can be seen in the model in {Ref.} \cite{KOIB-PRE-2004-2}, while it is not in the SA model at least under ${\it \Delta}p\!=\!0$.

\begin{figure}[hbt]
\vspace{3cm}
\centering
\includegraphics[width=12.5cm]{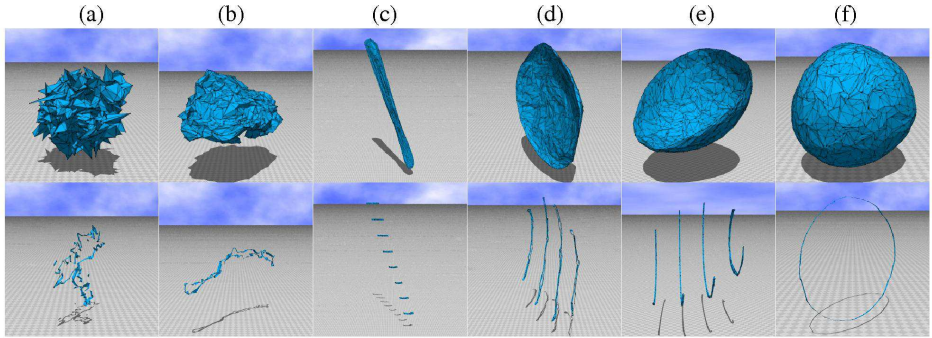}  
\caption{The snapshots of FC surfaces and the surface sections of size $N\!=\!1442$ obtained under ${\it \Delta}p\!=\!-0.5$ at (a) $\alpha\!=\!0$ (collapsed), (b) $\alpha\!=\!100$ (collapsed), (c) $\alpha\!=\!1.5\!\times\! 10^4$ (collapsed tubular), (d) $\alpha\!=\!2\!\times\! 10^4$ (cup like), (e) $\alpha\!=\!3\!\times\! 10^5$ (cup like), and (f) $\alpha\!=\!4\!\times\! 10^5$ (smooth spherical).} 
\label{fig-3}
\end{figure}
Snapshots of the FC surfaces and the surface sections are shown in Figs. \ref{fig-3}(a)--\ref{fig-3}(f), where a negative pressure ${\it \Delta}p\!=\!-0.5$ is assumed. The snapshots are slightly different from those at ${\it \Delta}p\!=\!0$ shown in Figs. \ref{fig-2}(a)--\ref{fig-2}(f). The snapshots in Figs. \ref{fig-3}(a) and \ref{fig-3}(b) indicate that the surfaces in the collapsed phase are almost crumpled. We see that the tubular surface in Fig. \ref{fig-3}(c) is also collapsed. The cup like surfaces in  Figs. \ref{fig-3}(d) and \ref{fig-3}(e) are new and typical of the condition ${\it \Delta}p\!=\!-0.5$, therefore, we call the new phase as the cup like phase. The smooth phase in  Fig. \ref{fig-3}(f) corresponds to the smooth phase in Fig. \ref{fig-2}(f) at ${\it \Delta}p\!=\!0$. The phase structure at $\alpha\!\to\!\infty$ is understood to be independent of ${\it \Delta}p$.   

 We also see that almost all parts of the surfaces in Figs. \ref{fig-2}(d) and \ref{fig-2}(e) consist of oblong triangles and are locally smooth along one specific direction and wrinkled along the direction vertical to the smooth direction. This is also expected in the linear phase shown in Fig. \ref{fig-3}(c) at ${\it \Delta}p\!=\!-0.5$. To the contrary, the surface in the wrinkled phase shown in Fig. \ref{fig-2}(a) consist of almost regular triangles and locally wrinkles along any directions. In the case of smooth phase in Fig. \ref{fig-2}(f), the surface is smooth along any directions. Thus, the surface is symmetric under the three-dimensional rotations both in the limit of $\alpha\!\to\!\infty$ and $\alpha\!\to\!0$, while the rotational symmetry is spontaneously broken at intermediate region of $\alpha$. This observation is independent of the two values of ${\it \Delta}p$. This symmetry breakdown or restoration is considered to be closely connected to the structural change of the constituent triangles; the symmetric surfaces are composed of almost regular triangles, while the non-symmetric surfaces are composed of oblong triangles. 

\begin{figure}[!h]
\centering
\includegraphics[width=10.0cm]{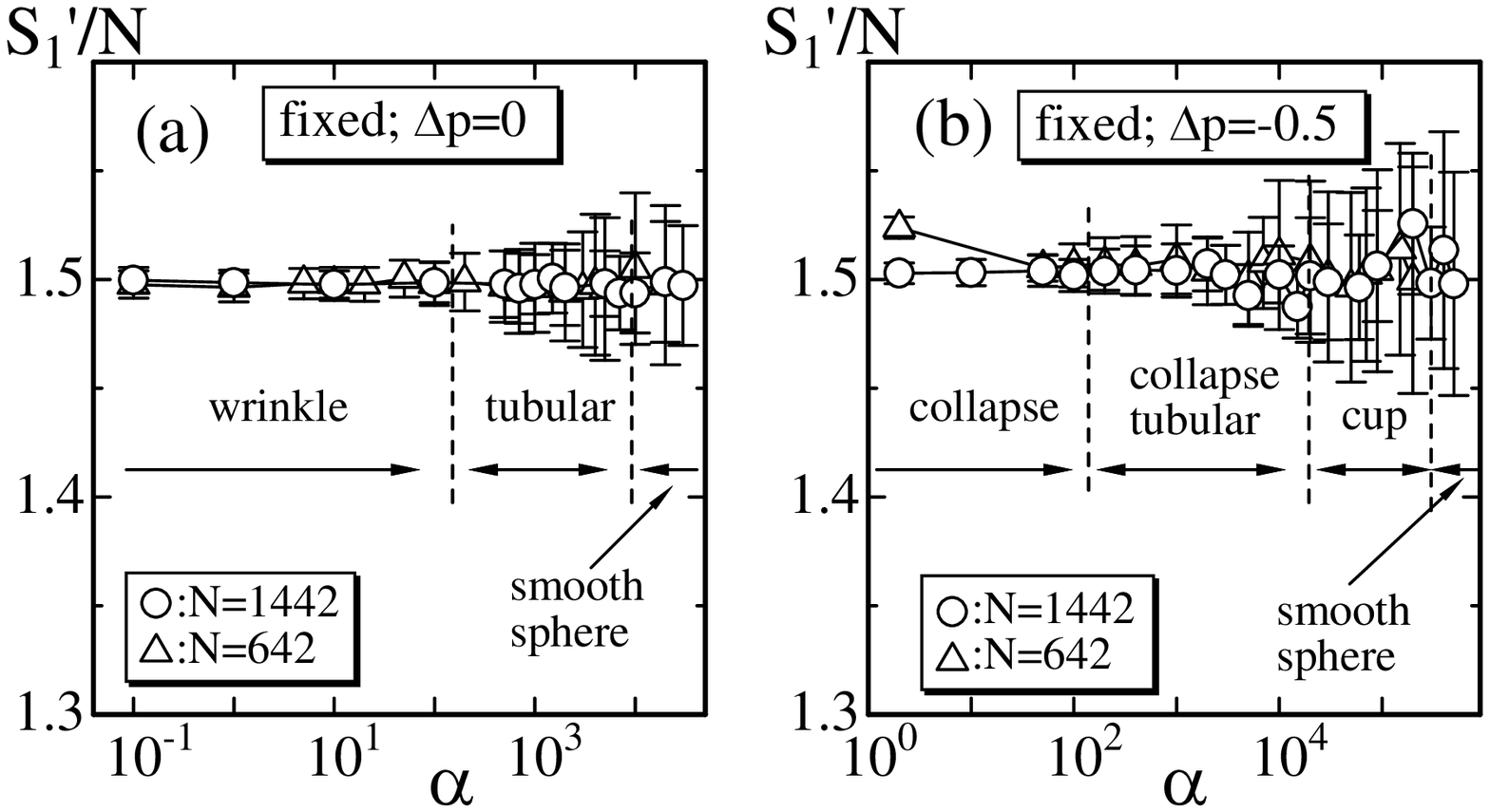}  
\caption{$[S_1\!-\!(3/2){\it \Delta}p\,V]/N$ vs. $\alpha$ under (a) ${\it \Delta}p\!=\!0$ and (b) ${\it \Delta}p\!=\!-0.5$. The error bars on the symbols denote the standard errors. The solid lines connecting the symbols are drawn as a guide to the eyes. } 
\label{fig-4}
\end{figure}
In the following presentations, we show how the shape transformation transitions and/or the SA interaction are reflected in the physical quantities including the Hausdorff dimension $H$ in the limit of $\alpha\!\to\! 0$. 

First of all, we show $[S_1\!-\!(3/2){\it \Delta}p\,V]/N$, denoted by $S_1^\prime/N$, in Figs. \ref{fig-4}(a) and \ref{fig-4}(b). Because of the scale invariant property of the partition function $Z_{\rm fix}$ of Eq. (\ref{Part-Func}), $S_1^\prime/N$ is expected to be $S_1^\prime/N\!=\!3/2$ at sufficiently large $N$. We see that all of the results are consistent with the prediction. This implies that the volume $V$ is well-defined and that the SA interaction is correctly implemented in the simulations.  We note that it is straightforward to prove that $S_1^\prime/N\!=\!3/2$ \cite{WHEATER-JP1994}. {As described in Section \ref {current_results},} the scale invariance of $Z$ is represented by $\partial Z(\lambda X)/\partial \lambda |_{\lambda=1} \!=\!0$.  {Because of the scale transformation {$X\to\lambda X$},} $S_1$ and $V$ change to $\lambda^2 S_1$ and $\lambda^3 V$ while $S_2$ and $U$ remain unchanged. Since the integration $\int \prod_i d X_i$ also changes to $\lambda^{3(N\!-\!1)}\int \prod_i d X_i$, then we have the relation  $S_1^\prime/N\!=\!3/2$ in the limit of $N\!\to\!\infty$. 

$S_1^\prime$ is identical with $S_1$, which is the total area of surface, in the case ${\it \Delta}p\!=\!0$, and therefore, the scale invariance implies that the surface area remains unchanged in the whole range of $\alpha$. To the contrary, the surface area $S_1$ discontinuously changes at the transition points in the case ${\it \Delta}p\!=\!-0.5$ at least, because $V$ discontinuously changes at the transitions as we will see below, while $S_1^\prime$ remains unchanged. This implies that the internal property of surface is significantly influenced by the external condition ${\it \Delta}p$.

\begin{figure}[!h]
\begin{center}
\includegraphics[width=10cm]{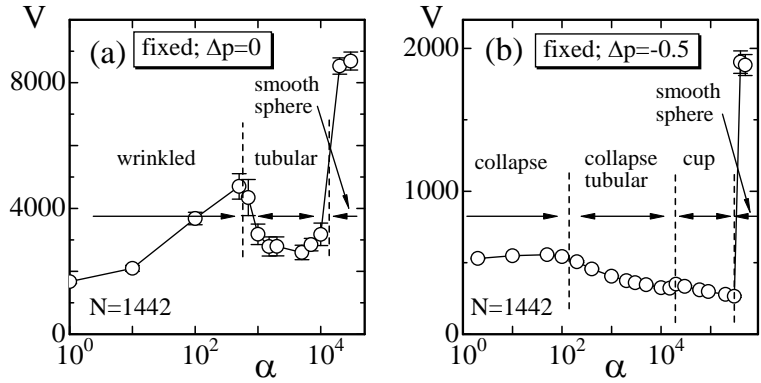}  
\caption{The volume $V$ vs. $\alpha$ under (a) ${\it \Delta}p\!=\!0$ and (b) ${\it \Delta}p\!=\!-0.5$. The vertical dashed lines denote the phase boundaries of the $N\!=\!1442$ surface. The solid lines connecting the symbols are drawn as a guide to the eyes.} 
\label{fig-5}
\end{center}
\end{figure}
The volume $V$ enclosed by the surface should be bounded below such that $V\geq 0$, which is satisfied only if the surface is self-avoiding. The model in this paper is strictly self-avoiding, and hence $V$ is expected to be well defined even when ${\it \Delta}p$ is large negative. Figures \ref{fig-5}(a) and \ref{fig-5}(b) show the dependence of $V$ on $\alpha$ under ${\it \Delta}p\!=\!0$ and ${\it \Delta}p\!=\!-0.5$, respectively. The vertical dashed lines in the figures represent the phase boundaries between two different phases just like in Fig. \ref{fig-4}. The name of the phases corresponds to the surface shape, which can be visualized as snapshots just like those in Figs. \ref{fig-2} and \ref{fig-3}.

The detailed informations such as the order of the transitions are not obtained. It is possible to perform the finite-size scaling analyses to see the order of the transitions by performing the simulations at the transition region more extensively, however, we confine ourselves of the phase structure in the wide range of $\alpha$ and, as a consequence, the order of the transitions is not fully examined. Thus, it remains unclear whether or not the smooth spherical phase and the tubular phase (or the cup like phase) are separated by a first-order transition, although the volume $V$ discontinuously changes at the phase boundary. We see that the volume $V$ in the collapsed phase is larger than that in the cup like phase under ${\it \Delta}p\!=\!-0.5$ at least, while $V$ at $\alpha\!\to\! 0$ is smaller than that in the tubular phase under ${\it \Delta}p\!=\!0$.

\begin{figure}[!h]
\centering
\includegraphics[width=10cm]{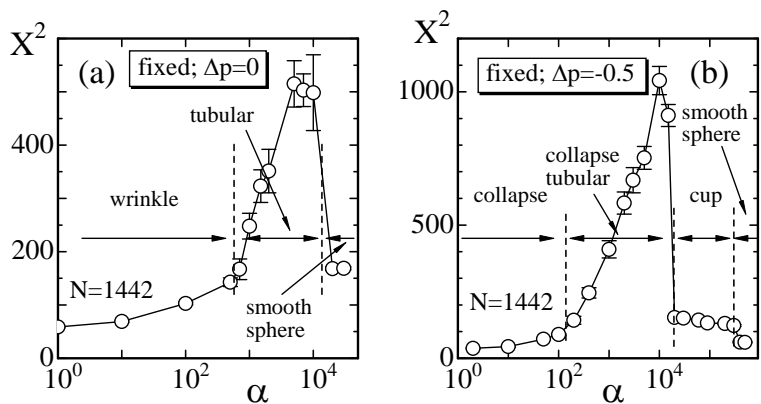}  
\caption{The mean square size $X^2$ vs. $\alpha$. The vertical dashed lines denote the phase boundaries of the $N\!=\!1442$ surface. } 
\label{fig-6}
\end{figure}
The mean square size $X^2$ is defined by
\begin{equation}
\label{X2}
X^2={1\over N} \sum_i \left(X_i-\bar X\right)^2, \quad \bar X={1\over N} \sum_i X_i,
\end{equation}
where $\bar X$ is the center of mass of the surface. The value of $X^2$ changes depending on the distribution of the vertices in ${\bf R}^3$, and hence $X^2$ as well as $V$ can reflect shape transformations. However, the quantity $X^2$ does not always show the same behavior against $\alpha$ as that of $V$. Figures \ref{fig-6}(a) and \ref{fig-6}(b) show $X^2$ vs. $\alpha$ under ${\it \Delta}p\!=\!0$ and ${\it \Delta}p\!=\!-0.5$. We see in Fig. \ref{fig-6}(a) that $X^2$ discontinuously changes at the phase boundary between the smooth spherical phase and the tubular phase. It is also easy to see from Fig. \ref{fig-6}(b) that $X^2$ discontinuously changes at the phase boundaries between the smooth spherical phase, the cup like phase, and the collapsed tubular phase. 
 
\begin{figure}[!h]
\centering
\includegraphics[width=12.5cm]{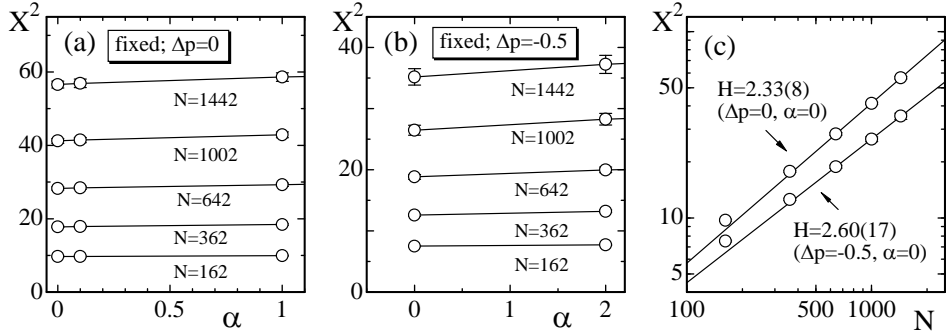}  
\caption{The mean square size $X^2$ vs. $\alpha$ at small $\alpha$ region under (a) ${\it \Delta}p\!=\!0$ and  (b) ${\it \Delta}p\!=\!-0.5$, and (c) $X^2$ vs. $N$ in a log-log scale obtained at $\alpha\!=\!0$ under ${\it \Delta}p\!=\!0$ and ${\it \Delta}p\!=\!-0.5$. The straight lines in (c) are drawn by fitting the largest three data points to Eq.(\ref{X2_scale}).    } 
\label{fig-7}
\end{figure}

\begin{table}[hbt]
\caption{ Hausdorff dimension $H$ obtained at $\alpha=0\sim 2$ under  ${\it \Delta}p\!=\!0$ and  ${\it \Delta}p\!=\!-0.5$. }
\label{table-1}
\begin{center}
 \begin{tabular}{ccccc}
              &  $\alpha=0$ & $\alpha=0.1$ & $\alpha=1$  & $\alpha=2$  \\
 \hline
  ${\it \Delta}p\!=\!0$  & $H=2.33\!\pm\!0.08$  &  $H=2.34\!\pm\!0.08$  & $H=2.33\!\pm\!0.08$ & - \\
  ${\it \Delta}p\!=\!-0.5$    & $H=2.60\!\pm\!0.17$  & - & - & $H=2.59\!\pm\!0.17$ \\
 \hline
 \end{tabular} 
\end{center}
\end{table}
Figures  \ref{fig-7}(a) and \ref{fig-7}(b) show $X^2$ obtained under ${\it \Delta}p\!=\!0$ and ${\it \Delta}p\!=\!-0.5$ at small $\alpha$ region. The Hausdorff dimension $H$ of the surface is defined by  
\begin{equation}
\label{X2_scale}
X^2\sim N^{2/H} \quad (N\to \infty).
\end{equation}
By fitting the data $X^2$ obtained at $\alpha\!=\!0$ to Eq. (\ref{X2_scale}), we draw straight lines in Fig. \ref{fig-7}(c), and the values of $H$ are shown in Table \ref{table-1} including $H$ at  $\alpha\!=\!0.1$ and $\alpha\!=\!1$ under ${\it \Delta}p\!=\!0$, and at $\alpha\!=\!2$ under  ${\it \Delta}p\!=\!-0.5$. The fitting is performed using the largest three data points under each condition of ${\it \Delta}p$.  We have  
$H\simeq 2.6$ at ${\it \Delta}p\!=\!-0.5$ and $H\simeq 2.33$ at ${\it \Delta}p\!=\!0$ in the limit of $\alpha\to 0$. The value of $H\simeq 2.33$ is compatible with the one $H\simeq 2.3$ in Ref. \cite{Gompper-Kroll-JPF1993}, while $H\simeq 2.6$ at ${\it \Delta}p\!=\!-0.5$ is slightly larger than the Flory estimate 2.5 and compatible with the fact that the surface is almost crumpled as we see in the snapshot in Fig. \ref{fig-3}(a). We should note that the value $H\simeq 2.3$ seems independent of the details of the model, the SA interaction, and the surface topology. However, the result $H\simeq 2.33$ is larger than the one $H\!=\!2.1(1)$ of Ref. \cite{BOWICK-TRAVESSET-EPJE2001}, thus it is also possible that $H$ depends on the model on the SA surfaces. 

We comment on the size effect of the results in Table \ref{table-1}. As mentioned above, the data obtained on the small sized surfaces, such as $N\!=\!162$ and $N\!=\!362$, were excluded from the fitting. By including the small two data in the fitting, we have $H\!=\!2.50(5)$ for ${\it \Delta}p\!=\!0$, $\alpha\!=\!0$ and  $H\!=\!2.93(7)$ for ${\it \Delta}p\!=\!-0.5$, $\alpha\!=\!0$. Both of $H$ are slightly larger than $H\!=\!2.33(8)$ and $H\!=\!2.60(17)$ shown in Table \ref{table-1}. Thus, the size effect is not negligible at least on the surfaces $N\!\leq\! 362$.

\begin{figure}[!h]
\centering
\includegraphics[width=10cm]{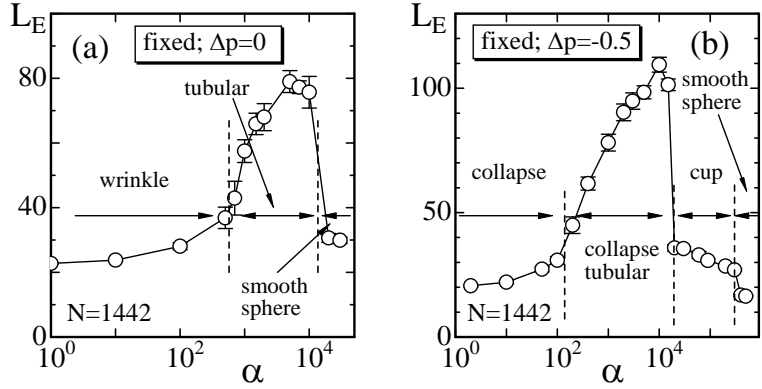}  
\caption{The maximum linear extension $L_E$ vs. $\alpha$. The dashed lines denote the phase boundaries of the $N\!=\!1442$ surface. } 
\label{fig-8}
\end{figure}
The surface size can also be reflected in the maximum linear extension $L_E$, which is defined by the maximum distance between two vertices on the surface:
\begin{equation}
\label{max-extentention}
L_E = {\rm Max} \{|X_i-X_j|\mid  (i,j=1,\cdots, N)\}, 
\end{equation}
where $X_i$ and $X_j$ are not always connected by a bond. The phase transition of shape transformation is also reflected in the structure of triangles; we see in the snapshots in Figs. \ref{fig-2} and \ref{fig-3} that the surface consists of almost regular triangles in the smooth spherical phase while it includes oblong triangles in the tubular phase, where $L_E$ expected to be very large. We expect that this structural change is reflected in $L_E$. 
Figures \ref{fig-8}(a) and \ref{fig-8}(b) show $L_E$ vs. $\alpha$ under ${\it \Delta}p\!=\!0$ and ${\it \Delta}p\!=\!-0.5$. The discontinuous change of $L_E$ at the phase boundaries shown in the figure implies that the phase transitions are accompanied by the structural change of surfaces. This structural change is typical of the Nambu-Goto surface model \cite{KOIB-EPJB-2007-3,KOIB-PRE-2004-2,KOIB-EPJB-2007-2}.

\begin{figure}[!h]
\centering
\includegraphics[width=10cm]{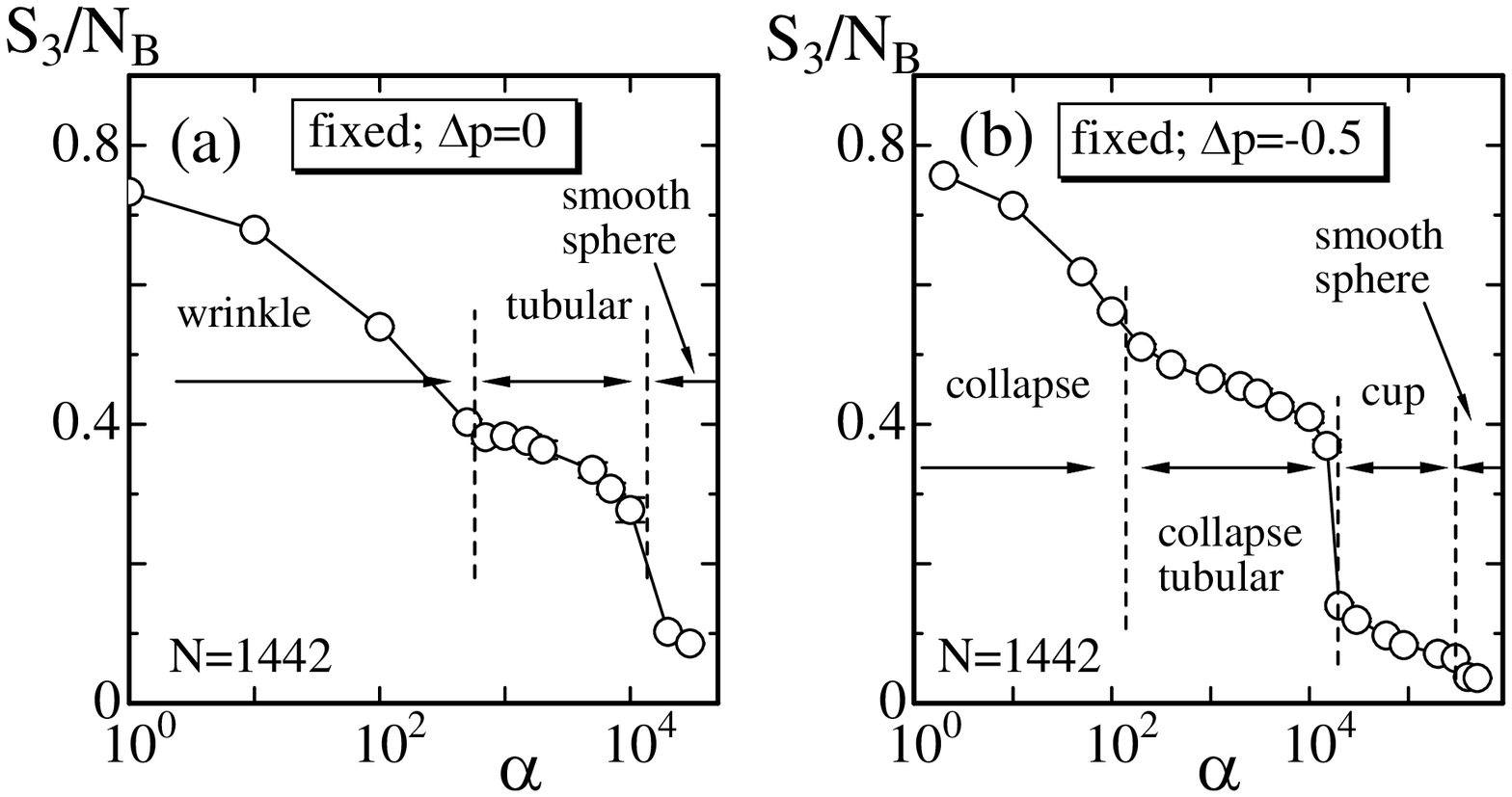}  
\caption{The two-dimensional bending energy $S_3/N_B$ vs. $\alpha$ under (a) ${\it \Delta}p\!=\!0$ and (b) ${\it \Delta}p\!=\!-0.5$.} 
\label{fig-9}
\end{figure}
The two-dimensional bending energy $S_3/N_B$ is shown in Figs. \ref{fig-9}(a) and \ref{fig-9}(b), where $S_3$ is defined by using a unit normal vector ${\bf n}_i$ of the triangle $i$ such that 
\begin{equation} 
\label{2-dim_bending}
S_3=\sum_{(ij)} (1-{\bf n}_i \cdot {\bf n}_j).
\end{equation} 
We write the two-dimensional bending energy as $S_3$ to distinguish it with the deficit angle energy $S_2$ in Eq. (\ref{Disc-Eneg}).

We see that $S_3/N_B$ discontinuously changes at the phase boundaries, where the physical quantities such as $V$, $X^2$ and $L_E$ discontinuously change. To the contrary, the deficit angle energy $S_2$ defined in Eq. (\ref{Disc-Eneg}), which is not shown in the figures, appears to vary almost smoothly in the whole range of $\alpha$. At the boundary between the smooth spherical phase and its neighboring phase, $S_2/N$ is expected to change discontinuously like the other physical quantities. However, the discontinuity is very small and it is almost invisible just as in the case of the self-intersecting model in {Ref.} \cite{KOIB-PRE-2004-2}. 
    
\section{Summary and Conclusion}\label{Conclusion}
We have numerically studied a self-avoiding (SA) surface model on fixed-connectivity (FC) triangulated lattices of sphere topology. The self-avoidance of the model in this paper is not identical to those of the well-known SA models; the ball spring model and the impenetrable plaquette (IP) model. However, the SA model in this paper belongs to the IP models, because the intersection of disjoint triangles are prohibited by the SA interaction. The phase structure of the FC model under ${\it \Delta}p\!=\!0$ is found to be almost identical to that of the phantom surface model in {Ref.} \cite{KOIB-PRE-2004-2} except for the evidence that the collapsed phase disappears from the SA model. Thus, the influence of the SA interaction on the phase structure is very small contrary to the expectation that the SA interaction can suppresses the multitude of phase transitions in the phantom surface model.  

To be more precise, the model in this paper is a Nambu-Goto surface model with a deficit angle energy. The SA interaction is defined such that all possible pairs of non-nearest neighbor triangles are prohibited from intersecting. Because the volume enclosed by the SA surface is well defined, the pressure term $-{\it \Delta}p\, V$ can be included in the Hamiltonian. The simulations are performed under ${\it \Delta}p\!=\!0$, and ${\it \Delta}p\!=\!-0.5$ on the FC surfaces, where ${\it \Delta}p\!=\!-0.5$ implies that the pressure inside the surface is lower than the pressure outside the surface.

Our observations on the FC surfaces are as follows: the smooth spherical phase, the tubular phase, and the collapsed phase can be seen under those two conditions of ${\it \Delta}p$, and the cup like phase is seen under ${\it \Delta}p\!=\!-0.5$. Thus, the phase structure of the model under ${\it \Delta}p\!=\!0$ is almost identical to that of the phantom surface model, although the collapsed phase is slightly different from each other; the collapsed surfaces are completely shrunk in the phantom surface model, while the SA surfaces are not completely shrunk at ${\it \Delta}p\!=\!0$ at least. The Hausdorff dimension $H\!=\!2.33(8)$, obtained at $\alpha\!=\!0$ under ${\it \Delta}p\!=\!0$, is independent of the curvature energy and is considered as the Hausdorff dimension of the Nambu-Goto SA surface. This result is consistent with the known result of $H\!\simeq\!2.3$ of the IP model in {Ref.} \cite{Gompper-Kroll-JPF1993}, where the model, the SA interaction, and the surface topology are different from those in this paper. In this sense, it is possible that the value $H\!\simeq\!2.3$ depends only on the self-avoidance, although the surface size of the simulation in {Ref.} \cite{Gompper-Kroll-JPF1993} is relatively smaller than those assumed in this paper. To the contrary, $H\!=\!2.33(8)$ is larger than the result $H\!=\!2.1(1)$ of {Ref.} \cite{BOWICK-TRAVESSET-EPJE2001}, and therefore, it is also possible that $H$ of the SA surface depends on the model. The SA surface models should be studied more extensively. 

It is also interesting to study whether or not the multitude of phases in the fluid surface models with cytoskeletal structures in Ref. \cite{KOIB-PRE-2007} is observed under a SA interaction. The SA interaction assumed in the model of this paper can also be assumed in those fluid surface models even when ${\it \Delta}p$ is negative. This remains to be a future study. 




\begin{thebibliography}{999}

\bibitem{NELSON-SMMS2004}
D. Nelson, in \textit{Statistical Mechanics of Membranes and Surfaces, Second Edition}, ed  D. Nelson, T. Piran, and S. Weinberg, (World Scientific, 2004, Singapore), p 1.  

\bibitem{Gompper-Schick-PTC-1994}
G. Gompper and M. Schick, \textit{Self-assembling amphiphilic systems}, In
\textit{Phase Transitions and Critical Phenomena 16}, ed Domb C and Lebowitz J L, (Academic Press, 1994, New-York) p 1. 

\bibitem{WIESE-PTCP19-2000}
K.J. Wiese, \textit{Phase Transitions and Critical Phenomena 19}, 
 ed C. Domb and J.L. Lebowitz, (Academic Press, 2000) p 253. 

\bibitem{Bowick-PREP2001}
M. Bowick and A. Travesset, 2001 Phys. Rep. {\bf 344} 255. 

\bibitem{SEIFERT-LECTURE2004}
U. Seifert, Fluid Vesicles, in \textit{Lecture Notes: Physics Meets Biology. From Soft Matter to Cell Biology.}, 35th Spring School, Institute of Solid State Research, Forschungszentrum J${\ddot {\rm u}}$lich (2004). 

\bibitem{GOMPPER-SMMS2004}
G. Gompper and D.M. Kroll, in \textit{Statistical Mechanics of Membranes and Surfaces, Second Edition}, ed  D. Nelson, T. Piran, and S. Weinberg, (World Scientific, 2004, Singapore), p 359.  

\bibitem{WHEATER-JP1994}
J.F. Wheater, 1994 J. Phys. A Math. Gen. {\bf 27} 3323. 


\bibitem{HELFRICH-1973}
W. Helfrich,  Z. Naturforsch {\bf 28}c,  693 (1973).

\bibitem{POLYAKOV-PLB1981}
A.M. Polyakov,  Phys. Lett. B {\bf 103}, 207, 211 (1981). 

\bibitem{POLYAKOV-NPB1986}
A.M.  Polyakov, Nucl. Phys. B {\bf 268}, 406 (1986). 

\bibitem{KLEINERT-PLB1986}
H. Kleinert,  Phys. Lett. B {\bf 174},   335 (1986). 

\bibitem{Peliti-Leibler-PRL1985}
L. Peliti and S. Leibler,  Phys. Rev. Lett. {\bf 54} (15),  1690 (1985). 

\bibitem{PKN-PRL1988}
M. Paczuski, M. Kardar, and D.R. Nelson,  Phys. Rev. Lett. {\bf 60},  2638 (1988). 

\bibitem{DavidGuitter-EPL1988}
F. David and E. Guitter,  Europhys. Lett. {\bf 5} (8), 709 (1988). 

\bibitem{Kownacki-Mouhanna-2009PRE}
J. -P. Kownacki and D. Mouhanna,  Phys. Rev. E {\bf 79},  040101 (R) (2009). 

\bibitem{BILLOIRE-DAVID-NPB1986}
A. Billoire and F. David,  Nucl. Phys. B {\bf 275} [FS17], 617 (1986). 

\bibitem{BKKM-NPB1986}
D.V. Boulatov, V.A. Kazakov, I.K. Kostov, and A.A. Migdal,  Nucl. Phys. B {\bf 275} [FS17],  641 (1986). 

\bibitem{KANTOR-NELSON-PRA1987}
Y. Kantor and D.R. Nelson,  Phys. Rev. A {\bf 36}, 4020 (1987). 

\bibitem{Baum-Ho-PRA1990}
A. Baumg${\ddot {\rm a}}$rtner and J.S. Ho,  Phys. Rev. A {\bf 41}, 5747 (1990). 

\bibitem{CATTERALL-NPBSUP1991}
S.M. Catterall, J.B. Kogut, and R.L.Renken,  Nucl. Phys. Proc. Suppl. B {\bf 99}A, 1 (1991). 

\bibitem{AMBJORN-NPB1993}
J. Ambjorn, A. Irback, J. Jurkiewicz, and B. Petersson,  Nucl. Phys. B {\bf 393}, 157 (1993). 

\bibitem{NISHIYAMA-PRE-2004}
Y. Nishiyama,  Phys. Rev. E {\bf 70}, 016101 (2004). 

\bibitem{KD-PRE2002}
J. -P. Kownacki  and H.T. Diep,  Phys. Rev. E {\bf 66}, 066105 (2002).

\bibitem{KOIB-PRE-2004}
H. Koibuchi, N. Kusano, A. Nidaira, K. Suzuki, and M. Yamada,  Phys. Rev. E {\bf 69},  066139 (2004). 

\bibitem{KOIB-PRE-2005}
H. Koibuchi and T. Kuwahata,   Phys. Rev. E {\bf 72},  026124 (2005).  

\bibitem{KOIB-NPB-2006}
I. Endo and H. Koibuchi,   Nucl. Phys. B {\bf 732} [FS], 426 (2006).  


\bibitem{KOIB-PRE-2007}
H. Koibuchi,  Phys. Rev. E {\bf 75}, 051115 (2007);  Phys. Rev. E {\bf 76}, 061105 (2007). 

\bibitem{KOIB-EPJB-2007-3}
H. Koibuchi,  Euro. Phys. J. B {\bf 59}, 405 (2007). 

\bibitem{KOIB-PRE-2004-2}
H. Koibuchi, Z. Sasaki, and K. Shinohara,  Phys. Rev. E {\bf 70}, 066144 (2004).  

\bibitem{KOIB-EPJB-2007-2}
Koibuchi H,  Euro. Phys. J. B {\bf 59}, 55 (2007). 

\bibitem{KANTOR-KARDAR-NELSON-PRL1986}
Y. Kantor, M. Karadar and D.R. Nelson,  Phys. Rev. Lett. {\bf 57}, 791 (1986). 

\bibitem{KANTOR-KARDAR-NELSON-PRA1987}
Y. Kantor, M. Karadar and D.R. Nelson,  Phys. Rev. A {\bf 35}, 3056 (1987). 

\bibitem{PLISCHKE-BOAL-PRA1988}
M. Plischke and Boal,  Phys. Rev. A {\bf 38}, 4943 (1988). 

\bibitem{Ho-Baum-EPL1990}
J. -S. Ho  and A. Baumg${\ddot {\rm a}}$rtner,  Europhys. Lett. {\bf 12},  295 (1990).

\bibitem{BAUM-JPIF1991}
A. Baumg${\ddot {\rm a}}$rtner,  J. Phys. I (France) {\bf 1}, 1549 (1991). 

\bibitem{BAUM-RENZ-EPL1991}
A. Baumg${\ddot {\rm a}}$rtner,  Europhys. Lett. {\bf 17}, 381 (1992). 

\bibitem{GREST-JPIF1991}
G. Grest,  J. Phys. I (France)  \textbf{1},  1695 (1991). 

\bibitem{Gompper-Kroll-JPF1993}
D.M. Kroll and G. Gompper,  J. Phys. France {\bf 3}, 1131 (1993). 

\bibitem{Gompper-Kroll-PRE1995}
G. Gompper and D.M. Kroll,  Phys. Rev. E {\bf 51}, 514 (1995). 

\bibitem{Munkel-Heermann-PRL1995}
C. M${\ddot {\rm u}}$nkel and D.M. Heermann,  Phys. Rev. Lett. {\bf 75}, 1666 (1995). 

\bibitem{BCTT-PRL2001}
M. Bowick, A. Cacciuto, G. Thorleifsson, and A. Travesset,  Phys. Rev. Lett. \textbf{87}, 148103 (2001).

\bibitem{BOWICK-TRAVESSET-EPJE2001}
M. Bowick, A. Cacciuto, G. Thorleifsson, and A. Travesset,  Euro. Phys. J. E \textbf{5}, 149 (2001). 

\bibitem{SWAMM-PRL-2010}
N. Stoop, F. K. Wittel, M. B. Amar, M. M. Muller, and H. J. Herrmann, Phys. Rev. Lett. \textbf{105}, 068101 (2010). 

\bibitem{ADF-NPB-1985}
J. Ambjorn, B. Durhuus, and J. Fr${\ddot {\rm o}}$hlich,  Nucl. Phys. B {\bf 257}, 433 (1985). 

\bibitem{Baillie-Johnston}
C.F. Baillie, and D.A. Johnston, 1993 Phys  Rev. D \textbf{48} 5025;  Phys. Rev. D \textbf{49}, 4139 (1994). 

\bibitem{BEJ}
C.F. Baillie, D. Espriu, and D.A. Johnston,  Phys. Lett. B \textbf{305}, 109 (1993). 

\bibitem{BIJJ}
C.F. Baillie, A. Irback, W. Janke, and D.A. Johnston,  Phys. Lett. B \textbf{325}, 45 (1994). 

\bibitem{KOIB-EPJB-2004}
H. Koibuchi, N. Kusano, A. Nidaira, Z. Sasaki, and T. Suzuki,  Euro. Phys. J. B {\bf 42}, 561 (2004). 

\bibitem{KOIB-PLA-2005}
M. Igawa, H. Koibuchi, and M. Yamada,  Phys. Lett. A {\bf 338}, 433 (2005). 

\bibitem{KOIB-PLA-2006-1}
I. Endo and H. Koibuchi,  Phys. Lett. A {\bf 350}, 11 (2006).

\bibitem{FDAVID-NPB-1985}
F. David, Nucl. Phys. B {\bf 257} [FS14], 543 (1985).

\bibitem{KOIB-EPJB-2007-1}
H. Koibuchi, Euro.  Phys. J. B {\bf 57}, 321 (2007).

\bibitem{BOAL-ZEIFERT-PRL1992}
D.H. Boal and U. Seifert, Phys. Rev. Lett. {\bf 69}, 3405 (1992).


\end{thebibliography}
\end{document}